\documentclass[11pt]{elsarticle}

\usepackage{graphics}
\usepackage[numbers,sort&compress]{natbib}

\usepackage{bm}
\newcommand{\ve}[1]{\bm{#1}}

\newcommand{\bu}{\ve{u}}

\newcommand{\bp}{\ve{p}}

\newcommand{\bmm}{\ve{m}}

\newcommand{\bq}{\ve{q}}
\newcommand{\bff}{\ve{f}}

\newcommand{\bx}{\ve{x}}

\newcommand{\bs}{\ve{s}}

\newcommand{\bRR}{\ve{\dot{R}}}

\newcommand{\bR}{\ve{R}}

\newcommand{\dd}{\text{d}}

\newcommand{\bphi}{\boldsymbol{\phi}}

\newcommand{\cI}{\mathcal{I}}

\newcommand{\cT}{\mathcal{T}}

\usepackage{tikz}
\usepackage{algpseudocode}
\usepackage{algorithm}
\usepackage{listings}
\usepackage{xcolor}
\usepackage{amsmath}  
\usepackage{booktabs} 
\usepackage{graphicx} 
\usepackage{inconsolata}
\usepackage{microtype}
\usepackage[margin=1in]{geometry}
\usepackage[parfill]{parskip}
\usepackage{hyperref}
\usepackage{url}
\usepackage{pgfplots}
\usepgfplotslibrary{groupplots}
\usepackage{ulem}
\usepackage{placeins}

\usepackage{caption}
\usepackage[]{siunitx}
\usepackage[nameinlink]{cleveref}
\crefname{lstlisting}{listing}{listings}
\Crefname{lstlisting}{Listing}{Listings}

\usepackage[dvipsnames]{xcolor}
\definecolor{lightblue}{rgb}{0.63, 0.74, 0.78}
\definecolor{seagreen}{rgb}{0.18, 0.42, 0.41}
\definecolor{orange}{rgb}{0.85, 0.55, 0.13}
\definecolor{silver}{rgb}{0.69, 0.67, 0.66}
\definecolor{rust}{rgb}{0.72, 0.26, 0.06}
\definecolor{purp}{RGB}{68, 14, 156}

\colorlet{lightrust}{rust!50!white}
\colorlet{lightorange}{orange!25!white}
\colorlet{lightlightblue}{lightblue}
\colorlet{lightsilver}{silver!30!white}
\colorlet{darkorange}{orange!75!black}
\colorlet{darksilver}{silver!65!black}
\colorlet{darklightblue}{lightblue!65!black}
\colorlet{darkrust}{rust!85!black}
\colorlet{darkseagreen}{seagreen!85!black}

\definecolor{RYB1}{RGB}{207, 37, 37}
\definecolor{RYB2}{RGB}{37, 91, 207}
\definecolor{RYB3}{RGB}{37, 207, 91}
\definecolor{RYB4}{RGB}{163,26,145}
\definecolor{RYB5}{RGB}{253, 180, 98}
\definecolor{RYB6}{RGB}{179, 222, 105}
\definecolor{RYB7}{RGB}{128, 177, 211}

\usepackage{hyperref}
\hypersetup{
  colorlinks=true,
}

\pgfplotscreateplotcyclelist{newcolors}{
{RYB1,every mark/.append style={fill=RYB1,mark size={2.5}},mark=*},
{RYB2,every mark/.append style={fill=RYB2},mark=square*},
{RYB3,every mark/.append style={fill=RYB3,mark size={3}},mark=triangle*},
{RYB4,every mark/.append style={fill=RYB4,mark size={3}},mark=diamond*},
{RYB5,every mark/.append style={fill=RYB5,mark size={3}},mark=pentagon*},
{RYB6,every mark/.append style={fill=RYB6,mark size={4}},mark=10-pointed star},
{RYB7,every mark/.append style={fill=RYB7},mark=*},
}

\tikzset{
    every node/.append style={font=\small},	
}

\pgfplotsset{
    	standard/.style={
    	compat=1.18,
        scale only axis,
        width=0.5\textwidth,
        enlarge x limits=0.05,
        enlarge y limits=0.05,
        max space between ticks=40,
        cycle list name=newcolors,
        every axis/.append style={font=\footnotesize},
        every legend/.append style={font=\footnotesize},
        every node/.append style={font=\footnotesize},	
	},
    cycle list name=newcolors,
    every axis/.append style={font=\footnotesize},
    every legend/.append style={font=\footnotesize},
    every node/.append style={font=\footnotesize},	
    compat=1.18,
}

\makeatletter
\def\ps@pprintTitle{%
   \let\@oddhead\@empty
   \let\@evenhead\@empty
   \let\@oddfoot\@empty
   \let\@evenfoot\@empty
}
\makeatother

\newcommand{\blfootnote}[1]{%
  \begingroup
  \renewcommand{\thefootnote}{}%
  \footnote{#1}%
  \addtocounter{footnote}{-1}%
  \endgroup
}

\begin{document}

\hypersetup{
  linkcolor=darkrust,
  citecolor=seagreen,
  urlcolor=darkrust,
  pdfauthor=author,
}

\begin{frontmatter}

\title{{\large\bf Hardware-Accelerated Phase-Averaging for Cavitating Bubbly Flows}}

\author[wpi]{Diego~Vaca-Revelo}
\author[gtcse]{Benjamin Wilfong}
\author[gtcse,gtae,gtme]{Spencer~H.~Bryngelson}
\author[wpi]{Aswin~Gnanaskandan}

\address[wpi]{Mechanical and Materials Engineering, Worcester Polytechnic Institute, Worcester, MA 01609, USA\vspace{-1ex}}
\address[gtcse]{School of Computational Science \& Engineering, Georgia Institute of Technology, Atlanta, GA 30332, USA\vspace{-1ex}}
\address[gtae]{Daniel Guggenheim School of Aerospace Engineering, Georgia Institute of Technology, Atlanta, GA 30332, USA\vspace{-1ex}}
\address[gtme]{George~W.~Woodruff School of Mechanical Engineering, Georgia Institute of Technology, Atlanta, GA 30332, USA}

\begin{abstract} 

We present a comprehensive validation, performance characterization, and scalability analysis of a hardware-accelerated phase-averaged multiscale solver designed to simulate acoustically driven dilute bubbly suspensions.
The carrier fluid is modeled using the compressible Navier--Stokes equations.
The dispersed phase is represented through two distinct subgrid formulations: a volume-averaged model that explicitly treats discrete bubbles within a Lagrangian framework, and an ensemble-averaged model that statistically represents the bubble population through a discretized distribution of bubble sizes.
For both models, the bubble dynamics are modeled via the Keller--Miksis equation.
For the GPU cases, we use OpenACC directives to offload computation to the GPUs.
The volume-averaged model is validated against the analytical Keller-Miksis solution and experimental measurements, showing excellent agreement with root-mean-squared errors of less than \SI{8}{\percent} for both single-bubble oscillation and collapse scenarios.
The ensemble-averaged model is validated by comparing it to volume-averaged simulations.
On an NCSA~Delta node with 4~NVIDIA A100 GPUs, we observe a speedup 16-fold compared to a 64-core AMD Milan CPU.
The ensemble-averaged model offers additional reductions in computational cost by solving a single set of averaged equations, rather than multiple stochastic realizations.
However, the volume-averaged model enables the interrogation of individual bubble dynamics, rather than the averaged statistics of the bubble dynamics.
Weak and strong scaling tests demonstrate good scalability across both CPU and GPU platforms.
These results show the proposed method is robust, accurate, and efficient for the multiscale simulation of acoustically driven dilute bubbly flows.

\end{abstract}


\end{frontmatter}

\blfootnote{Code available at \url{https://github.com/MFlowCode/MFC}.}

\section{Introduction} \label{sec:intro}

Acoustic cavitation refers to the process in which sound waves induce rapid pressure fluctuations within a liquid, causing the formation, growth, shrinkage, and eventual collapse of small gas-filled bubbles~\cite{neppiras1980, yasui2017}.
These bubbles oscillate in response to the alternating phases of compression and rarefaction in the acoustic field.
Under suitable conditions, the oscillations can become unstable, leading to inertial collapse that concentrates energy into a small region.
This collapse can generate extreme local temperatures, high-pressure shock waves, and high-velocity microjets~\cite{flint1991, lauterborn2010, beig2018}.
Acoustic cavitation has been widely applied in fields such as biomedicine~\cite{ikeda2016, kooiman2020, coussios2008applications} and engineering~\cite{bang2010materials, luo2014biomass, suslick1999chemical}, where the unique properties of bubble dynamics are exploited for various purposes.

While acoustic cavitation can occur in a wide range of bubbly environments, a particularly important case is that of dilute bubbly suspensions.
In such systems, small gas bubbles are dispersed within a liquid, with a low volume fraction (typically less than $0.01$) of bubbles.
The bubble distribution, bubble size, acoustic bubble-bubble interaction, and the surrounding liquid typically influence the behavior of these suspensions under the influence of an acoustic wave.
Even at low concentrations, these small bubbles strongly influence sound propagation, introducing dispersion and attenuation which becomes prominent when the bubbles oscillate near their resonance frequencies~\cite{voronin2003, leroy2008}.
These characteristics make dilute bubbly suspensions a versatile medium for both theoretical and experimental validation of acoustic models. Applications of dilute bubbly suspensions span multiple disciplines.
In medical ultrasound, microbubble suspensions are used as contrast agents to enhance diagnostic imaging and as carriers for targeted drug delivery, where controlled cavitation can release therapeutic compounds at precise locations~\cite{kooiman2020, coussios2008applications}.
It is also central to shock wave lithotripsy, a non-invasive medical procedure that uses acoustic cavitation-induced shock waves to break down kidney stones~\cite{ikeda2016}.
Furthermore, microbubble-enhanced high-intensity focused ultrasound therapy leverages acoustic cavitation to enhance the therapeutic effects of ultrasound in medical treatments, such as tumor ablation~\citep{kaneko2005, kajiyama2010, chung2012, juang2023controlled}.
It has also been utilized in materials processing, including in the enhancement of chemical reactions through sonochemistry~\cite{bang2010materials, luo2014biomass}.
Industrial processes make use of their cleaning capabilities, emulsification potential, and effectiveness in food processing, while environmental engineering applications include wastewater treatment, where enhanced aeration and oxidation accelerate contaminant breakdown~\cite{canselier2002, krasulya2016, dular2016}.
Therefore, elucidating a better understanding of the dynamics of gas bubbles dispersed in a liquid medium under the influence of acoustic waves is of vital importance.

The length scales in dilute bubbly flows span centimeter-scale acoustic wave propagation and micrometer-scale bubbles, and fully resolving both would require an impractically fine mesh across the entire domain.
Subgrid bubble models are therefore essential for representing such multiscale bubbly systems efficiently.
In this framework, the acoustic field is resolved on a relatively coarse Eulerian grid, while the unresolved microbubbles are modeled at the subgrid scale.
Two phase-averaged subgrid-scale approaches are commonly employed: ensemble averaging (Euler--Euler) and volume averaging (Euler--Lagrange).
In the ensemble-averaged approach, the aim is to capture the mean collective response of a bubble population characterized by a known probability distribution of radii.
This method assumes the presence of many stochastically distributed bubbles within each computational cell and evaluates the statistically averaged mixture dynamics.
Individual bubbles are not explicitly resolved, though this approach retains the essential acoustic and dynamical effects of the unresolved bubbles, offering good computational efficiency~\cite{zhang1994,ando2011}.
Volume-averaging employs a Lagrangian strategy, where individual bubbles are treated as discrete entities~\cite{maeda2018eulerian, gnanaskandan2019modeling}, and their volumetric oscillations are typically modeled using one of the standard bubble dynamics equations. While this approach retains the individual size dynamics of each bubble, it is usually expensive to obtain the mean behavior of a bubble population due to the need for several independent ensembles of bubble distributions within a cloud.
The computationally advantageous approach depends on the number of bubbles within the bubbly suspension and the available computational resources.
Typically, the ensemble-averaged models are preferable for a large number of bubbles, and volume-averaged models are preferred when representing individual bubble dynamics is important~\cite{bryngelson2019quantitative}.

Although these models are well-established, improving their computational performance remains a priority for enabling large-scale parametric studies that can close knowledge gaps in cavitation physics.
A common acceleration strategy is distributed memory parallelization via MPI~\cite{snir1998mpi}.
With a purely grid-based decomposition, each CPU core performs separate calculations, and the MPI protocol exchanges data with its neighboring cores.
This strategy is effective when the computational workload is nearly evenly distributed, such as when solving the compressible Navier--Stokes equations in a finite volume framework, where each subdomain typically requires a similar amount of computations if the number of cells is distributed evenly.
Simulations involving subgrid bubbles, on the other hand, can introduce load balancing challenges~\cite{bohme2014characterizing}.
Load imbalance occurs when bubbles are localized in specific regions of the domain, resulting in an uneven computational workload across processors.
This issue may appear in both the phase-averaged subgrid bubble models, being more prejudicial for the volume-averaged model, since for the discrete phase, the number of equations being solved is proportional to the number of bubbles present.
Subdomains with high bubble concentrations require significantly more processing time, while others with fewer or no bubbles finish their computations sooner.
The underutilized processors must then wait for the heavily loaded ones to complete before MPI synchronization can proceed.
This unproductive time reduces parallel efficiency and slows the overall simulation, introducing an opportunity to improve existing solvers.
The present work addresses this by optimizing solver performance and mitigating the effects of load imbalance.

Another well-established approach for accelerating CFD solvers is the use of offloading strategies, particularly to modern graphics processing units (GPUs)~\cite{wang2010adaptive, salvadore2013gpu, sweet2018, radhakrishnan2024}.
GPUs can offer substantial performance gains; for example, \citet{sweet2018} reported speedups of up to 14x in simulations of particle-laden turbulent flows compared to CPU-only implementations.
\citet{piscaglia2023} used OpenFOAM and offloaded computationally intensive calculations to GPUs and reported a 10-fold speedup.
The acceleration reported in the work of \citet{jespersen2010} goes from 2.5x to 3.0x when comparing a GPU against a single CPU.
Nevertheless, GPU acceleration may be constrained by limited memory and increased communication overhead.
To overcome these issues, \citet{radhakrishnan2024} developed a directive-based offloading strategy for multiphase compressible flow solvers, achieving high memory reuse and efficient computation.
This strategy was implemented in the open-source CFD solver Multi-Component Flow Code (MFC)~\cite{bryngelson2021mfc,wilfong2025}, where a speedup of 40x was reported on a single node using NVIDIA V100 GPUs over IBM POWER9 CPUs.
Building on this foundation, the present work introduces a GPU-based hardware acceleration strategy for phase-averaged subgrid models to enhance computational efficiency.

This work explores the computational speedup achievable with the state-of-the-art subgrid bubble models on CPU and GPU architectures, identifies the conditions under which each phase-averaged formulation is most computationally efficient, and evaluates how effectively modern GPU hardware can mitigate load imbalance arising from localized bubble clusters.
To address these questions, we develop and assess a GPU-accelerated framework for both volume- and ensemble-averaged subgrid bubble models, quantify the speedup relative to CPU-based execution, and examine the performance of each approach across a range of bubble populations.
The analysis identifies the scenarios in which each model offers the greatest computational advantage and shows how GPU acceleration enables simulations over a broader range of parameters and large-scale systems that would not be feasible with CPU-only computations.

In the following sections, we present the computational methodology in detail, validate the solver against several representative test cases, highlight the substantial performance gains achieved with GPU-based computations compared to CPU-only simulations, and demonstrate code scalability through strong and weak scaling studies on both CPU and GPU architectures.

\section{Governing equations} \label{sec:governing_eqns}

At the macroscale, acoustic cavitation problems are resolved using a fixed-grid Eulerian framework, while the influence of unresolved bubbles is incorporated via phase-averaged subgrid-scale models.
In dilute bubbly suspensions, the propagation of acoustic waves can be described by a fully compressible continuum model of the liquid--bubble mixture.
Any mixture property, denoted by $(\cdot)$ is defined as $(\cdot)=(1-\alpha)(\cdot)_l+\alpha(\cdot)_g$, where $\alpha$ is the volume fraction of the gas contributed by the bubbles, and the subscripts $l$ and $g$ denote the liquid and gas phases, respectively.
The governing conservation equations for mass, momentum, and energy take the following form:
\begin{equation} \label{eqn:euler_transport}
    \begin{aligned}
        \frac{\partial \rho}{\partial t}+\nabla \cdot(\rho \bu) &= 0, \\
        \frac{\partial(\rho \bu)}{\partial t}+\nabla \cdot(\rho \bu \otimes \bu + p \cI - \cT ) &= \ve{0}, \\
        \frac{\partial E}{\partial t}+\nabla \cdot\left[(E+p) \bu-\cT \cdot \bu\right] &= 0,
    \end{aligned}
\end{equation}
where $\rho$ is the density, $\bu$ is the velocity vector, $p$ is the pressure, and $E$ is the total energy.
The term $\cT$ denotes the effective viscous stress tensor of the mixture.
In dilute suspensions, the characteristic low void fraction, up to O($10^{-2}$), allows us to treat the liquid density as significantly greater than that of the gas $\rho_l >> \rho_g$ \cite{maeda2018eulerian, ando2011}.  Then, we can approximate the mixture density to be $\rho \approx (1-\alpha) \rho_l$. Additionally, we assume zero slip velocity between the phases $\bu \approx \bu_l =\bu_g$. Thus, there is effectively no momentum transfer across the gas–liquid interface, allowing us to approximate the effective viscous stress as that of the continuous phase:

\begin{equation} \label{eqn:stressTensor}
    \cT \equiv \cT_l = \mu_l\left(\nabla \bu+\nabla \bu^{\top}-\tfrac{2}{3}(\nabla\cdot \bu) \cI \right)
\end{equation}
where $\mu_l$ is the liquid viscosity.

To account for the presence of the bubbles and their interaction with the surrounding liquid, we use the volume- and ensemble-averaged models.
Each of them is derived from a distinct set of governing equations, based on physically justified simplifying assumptions.
Detailed formulations are provided in the subsequent subsections of the manuscript.
Both phase-averaged models assume that the bubbles are spherical and sufficiently separated so that collisions, bouncing, and coalescence can be ignored.
Additionally, bubble--bubble interaction occurs only through their effect on the liquid-bubble mixture.

In both subgrid models, the volumetric oscillations of the bubbles in response to pressure variations in the surrounding liquid are described by the Keller--Miksis equation, which incorporates the liquid's compressibility effect, and is given by:
\begin{equation}
    \begin{aligned} \label{eqn:keller_miksis}
    \left(R\left(1-\frac{\dot{R}}{c}\right)\right) \ddot{R} + \frac{3}{2} \dot{R}^2 \left(1-\frac{\dot{R}}{3 c}\right) &= \frac{p_{bw}-p_{\infty}}{\rho}\left(1+\frac{\dot{R}}{c}\right)+\frac{R \dot{p_{bw}}}{\rho c}, \\
    p_{bw} &= p_{b}-\frac{4 \mu_l \dot{R}}{R}-\frac{2 \sigma}{R},
    \end{aligned}
\end{equation}
where $R$, $\dot{R}$, and $\ddot{R}$ are the radius, interface velocity, and interface acceleration of the bubble.
The term $p_{bw}$ is the pressure at the bubble wall, $p_{b}$ is the pressure inside the bubble, and $p_\infty$ is the pressure that forces the radial oscillations of the bubble. $\mu_l$ is the dynamic viscosity of the background medium, $\sigma$ is the surface tension, and $c$ is the liquid's speed of sound.
We assume that the bubble contains both non-condensable gas and vapor, and we adopt the reduced-order models with constant heat and mass transfer coefficients at the bubble wall, as described by \citet{preston2007}.
These models account for the effects of vapor and heat diffusion through the interface.
Here, the vapor mass transfer rate is given by
\begin{equation}
    \dot{m}_v = \frac{\mathcal{D} \rho_{bw}}{1 - \chi_{vw}} \frac{\partial \chi_{vw}}{\partial r}\Bigr|_{\substack{r=R}}, \label{eqn:massVaporRate}
\end{equation}
where $\chi_v$ is the vapor mass fraction, $\mathcal{D}$ is the binary diffusion coefficient, and subscript $w$ denotes properties at the bubble wall.
The internal pressure $p_{b}$ evolves following the model of \citet{ando2011} as follows:
\begin{equation}
    \dot{p}_b = \frac{3 \gamma_b}{R} \left(- \dot{R} p_b + \mathcal{R}_v T_{bw} \dot{m_v} + \frac{\gamma_b - 1}{\gamma_b} k_{bw} \frac{\partial T}{\partial r}\Bigr|_{\substack{r=R}} \right),
\end{equation}
where $\gamma_b$ is the specific-heat ratio of the bubble contents, $\mathcal{R}_v$ is the gas constant of vapor, $T_{bw}$ is the bubble-wall temperature and $k_{bw}$ is the thermal conductivity for the bubble contents.

In dilute bubbly suspensions, external acoustic perturbations are necessary to excite bubble oscillations and sustain cavitation dynamics.
In this work, we use a source-term approach to generate one-way acoustic waves~\citep{maeda2017source}.
This method enables the injection of unidirectional acoustic waves from an arbitrarily shaped source surface by introducing appropriate forcing terms into the mass, momentum, and energy equations on that surface within the computational domain.

\begin{figure}[t]
    \centering
    \includegraphics[scale=1]{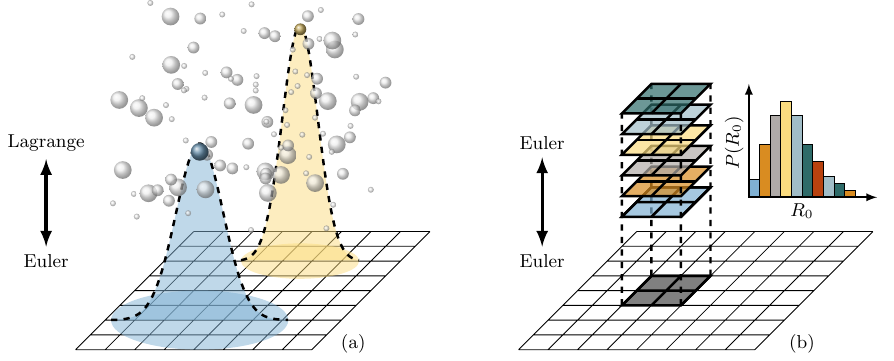}
    \caption{Schematics of the (a) volume-averaged and (b) ensemble-averaged subgrid bubble models.}
    \label{fig:el_ee_schematic}
\end{figure}

\subsection{Volume-averaged (EL) subgrid model}

This model is formulated within an Euler--Lagrange (EL) framework, where the bubbles are treated as discrete entities in three-dimensional space and their interaction with the carrier fluid is represented through a two-way coupling strategy.
Our volume-averaged model (or EL model) follows the formulation of \citet{maeda2018eulerian}.
In cavitation problems, a bubble's behavior is influenced not only by its radial oscillations but also by its translational motion.
However, the formulation of \citet{maeda2018eulerian} simplifies the problem by assuming that bubbles remain fixed in space.
As a result, bubble translation, which would require solving an additional equation of motion, is neglected.
This assumption provides a reasonable approximation because the timescale is typically much longer than that of the radial dynamics.
Under moderate acoustic forcing and in the absence of additional perturbations, the induced liquid velocities remain small, making bubble motion negligible compared to the rapid radial oscillations of the bubble.

The two-way Euler--Lagrange coupling is established as follows: the far-field pressure driving the bubble oscillations, $p_{\infty}$, is obtained from the Eulerian pressure field, while the bubbles' influence is transferred back by smearing their effect on the background grid.
Applying the model assumptions to \cref{eqn:euler_transport}, the following inhomogeneous hyperbolic system is derived:
\begin{equation}
    \begin{aligned}
    \frac{\partial \rho_l}{\partial t}+\nabla \cdot\left(\rho_l \bu_l \right) & =\frac{\rho_l}{1-\alpha}\left[\frac{\partial \alpha}{\partial t}+ \bu_l \cdot \nabla \alpha\right], \\
    \frac{\partial\left(\rho_l \bu_l\right)}{\partial t}+\nabla \cdot\left(\rho_l \bu_l \otimes \bu_l+p_l \cI-\cT_l\right) & =\frac{\rho_l \bu_l}{1-\alpha}\left[\frac{\partial \alpha}{\partial t}+\bu_l \cdot \nabla \alpha\right]-\frac{\alpha \nabla \cdot\left(p_l \cI-\cT_l\right)}{1-\alpha}, \\
    \frac{\partial E_l}{\partial t}+\nabla \cdot\left(\left(E_l+p\right) \bu_l-\cT_l \cdot \bu_l\right) & =\frac{E_l}{1-\alpha}\left[\frac{\partial \alpha}{\partial t}+\bu_l \cdot \nabla \alpha\right]-\frac{\alpha \nabla \cdot\left(p \bu_l-\cT_l \cdot \bu_l\right)}{1-\alpha}.
    \end{aligned}
\end{equation}
The left-hand side corresponds to the conservation equations of the liquid phase, while the right-hand side represents source terms that incorporate the effects of the bubbles.
The liquid pressure $p_l$ follows the stiffened-gas equation of state
\begin{equation} \label{eqn:eos_liquid}
    p_{l} = \left( {\gamma_{l} - 1} \right) \rho_l \varepsilon_l - \gamma_l\pi_{\infty, l},
\end{equation}
where $\varepsilon_l$ is the liquid's internal energy, and $\gamma_l$ and $\pi_{\infty, l}$ are the specific heat ratio and the stiffness of the liquid, respectively.

The instantaneous bubble sizes are communicated to the background flow solver through the local void fraction, $\alpha$.
This coupling is achieved by computing an effective void fraction that distributes each bubble's volume contribution to the surrounding computational cells, as illustrated in \cref{fig:el_ee_schematic}~(a).
At the bubble center $\bx_n$ of bubble $n$, the volume is smeared into the continuous void fraction field by a regularization kernel $\delta$, giving
\begin{gather}
    \alpha(\bx_n) =
        \sum_{n=1}^N V_n \delta = \sum_{n=1}^N \left( \frac{4}{3} \pi R_n^3 \right) \delta,
\end{gather}
where $N$ is the total number of bubbles and $V_n$ is the volume.
We use a continuous, second-order truncated Gaussian kernel:
\begin{equation}\label{eqn:kernel3D}
    \delta(d_n,h) = 
        \begin{cases}
            \frac{1}{h^3(2 \pi)^{3 / 2}} e^{-\frac{d_n^2}{2 h^2}} & 0 \leq \frac{d_n}{h}<3, \\ 
            0, & 3 \leq \frac{d_n}{h},
    \end{cases}
\end{equation}
where $h$ is the kernel width and $d_n=\left|\bx-\bx_n\right|$ is the distance to the bubble center.
The evolution of $\alpha$ follows as:
\begin{equation}
    \frac{\partial \alpha(\bx)}{\partial t} = 
        \sum_{n=1}^N \frac{\partial V_n}{\partial t} \delta+\sum_{n=1}^N V_n \frac{\partial \delta}{\partial t},
\end{equation}
with
\begin{equation}
    \frac{\partial V_n}{\partial t} = 4 \pi R_n^2 \dot{R}_n,
    \quad 
    \frac{\partial \delta}{\partial t} = -\bu_l \cdot \nabla \delta.
\end{equation}

\citet{maeda2018eulerian} impose a constraint on the kernel support width $h$ to ensure that the model resolves the small-scale dynamics inside the bubble cloud:
\begin{equation} \label{eqn:model_inequality}
     \left\{ \begin{array}{c} R_b \\ \Delta \end{array} \right\} \le h < L_b,
\end{equation}
where $R_b$ is the characteristic bubble radius, $\Delta$ is the Eulerian grid spacing, and $L_b$ is the characteristic inter-bubble distance.
This condition prevents kernel supports from overlapping and ensures that the physics at the inter-bubble scale are accurately captured.
Because the maximum bubble radius $R_b$ is not known a priori, the solver dynamically adjusts $h$ to satisfy \cref{eqn:model_inequality}.
As long as this condition holds, the model is valid even when bubbles grow larger than the grid size.
Accordingly, the solver sets $h=\Delta$ when $R_b<\Delta$, and $h=R_b$ otherwise.
A more detailed description of the volume-averaged model can be found in~\cite{maeda2018eulerian}.

\subsection{Ensemble-averaged (EE) subgrid model}

This model can be viewed as an Euler--Euler (called EE) formulation, as illustrated in \cref{fig:el_ee_schematic}~(b).
Here, instead of solving for the dynamics of individual bubbles, it evaluates the statistically-averaged mixture dynamics by assuming a large number of stochastically scattered bubbles dispersed within each computational grid cell.
Our ensemble-averaged model follows the description of \citet{zhang1994} and \citet{bryngelson2019quantitative}. 

The equilibrium radii of the bubbles, $\bR_{0}$, are assumed to follow a log-normal distribution, which is further discretized using $N_{\mathrm{bin}}$ number of bins.
The bubble population is represented statistically through the variables $\bR$, $\bRR$, $\bp_{b}$, and $\bmm_{v}$, corresponding to the instantaneous bubble radii, interface velocities, bubble pressures, and vapor mass.
Each of these variables contains $N_{\mathrm{bin}}$ components.
The mixture-averaged pressure in \cref{eqn:euler_transport} is given by
\begin{equation}
    p = (1-\alpha) p_\ell +
	\alpha  \left(
		\frac{\overline{\bR^3 \bp_{bw} }}{\overline{\bR^3}} - \rho \frac{ \overline{ \bR^3 \dot{\bR}^2 }}{ \overline{\bR^3} }
	\right),
    \label{e:pressure}
\end{equation}
where $\bp_{bw}$ is the associated bubble wall pressure defined in \cref{eqn:keller_miksis}.
The liquid pressure $p_\ell$ is computed using the stiffened-gas equation of state given in \cref{eqn:eos_liquid}.
The bubble number density per unit volume $n_\mathrm{bub.}(\bx,t)$ is conserved as
\begin{equation}
    \frac{\partial n_\mathrm{bub.} }{\partial t } + \nabla \cdot ( n_\mathrm{bub.} \bu ) = 0.
    \label{e:consn}
\end{equation}
For the spherical bubbles considered here, $n_\mathrm{bub.}$ is related to the void fraction, $\alpha$, via the conservation of the number density function:
\begin{equation}
    \alpha(\bx,t) = \frac{4}{3} \pi { \overline{\bR^3} } n_\mathrm{bub.}(\bx,t),
    \label{e:ndf}
\end{equation}
and so the void fraction $\alpha(\bx,t)$ transports as
\begin{equation}
    \frac{\partial \alpha }{\partial t } + 
        \nabla \cdot (\alpha \bu) =
	3 \alpha \frac{ \overline{\bR^2 \bRR }}{ \overline{\bR^3} },
    \label{e:alpha}
\end{equation}
where the right-hand side represents the change in void fraction resulting from the growth and collapse of bubbles.
The bubble dynamics are evaluated as
\begin{equation}
    \frac{\partial n_\mathrm{bub.} \bphi }{\partial t } + 
        \nabla \cdot (n_\mathrm{bub.} \bphi \bu) =
	n_\mathrm{bub.} \dot{\bphi},
\end{equation}
where $\bphi=\{\bR, \bRR, \bp_{b}, \bmm_{v}\}$ contains the bubble dynamic variables.
The over-barred terms in the above equations denote averages computed across the bubble dispersion, confining all $N_{\mathrm{bin}}$ bubble groups in each control volume.
All over-barred terms require a numerical closure, which is accomplished by distributing the bubble equilibrium radii $R_o$ in $N_\mathrm{bin}$ bins with a log-normal probability distribution function (PDF).
The integration of these terms follows from Simpson's rule, though more advanced techniques are available in the limit of small bubble oscillation amplitude~\citep{bryngelson_levin25,sinha24neural}.
The Euler--Euler ensemble averaging technique can be readily extended to a population balance formulation that accounts for distributions in all independent bubble coordinates, $R$, $\dot{R}$, and $R_o$~\citep{bryngelson2023conditional,charalampopoulos21}.
More details of the ensemble-averaged model can be found in~\cite{zhang1994,colonius2008statistical}.

\section{Numerical method} \label{sec:numerical_method}

The conservative form of the set of governing equations \cref{eqn:euler_transport} can be generalized as:
\begin{equation} \label{eqn:generic_transport}
    \frac{\partial \bq}{\partial t}+\nabla \cdot \bff(\bq) = \bs,
\end{equation}
where, $\bq$ is the vector of conservative variables, $\bff(\bq)$ are the fluxes, and $\bs$ contains any source terms.
To numerically solve \cref{eqn:generic_transport}, MFC employs the finite volume method.
This equation can be spatially discretized in a Cartesian framework as
\begin{equation}
    \frac{\partial \bq}{\partial t}+\frac{\partial \bff^\chi\left(\bq\right)}{\partial x}+\frac{\partial \bff^y\left(\bq\right)}{\partial y}+\frac{\partial \bff^z\left(\bq\right)}{\partial z}=\bs,
\end{equation}
where $\bff^x, \bff^y$ and $\bff^z$ are vectors of fluxes in the $x$, $y$, and $z$ directions.
The above equation is integrated over each finite volume 
grid cell $(i,j,k)$ in the three coordinate directions.
The resultant equation in semi-discrete form is
\begin{equation}
    \begin{aligned}
        \frac{\dd \bq_{i, j, k}}{\dd t} = 
        \frac{1}{\Delta x_i}
        \left[\bff_{i-1 / 2, j, k}^x-\bff_{i+1 / 2, j, k}^x\right] + 
        \frac{1}{\Delta y_j}
        \left[\bff_{i, j-1 / 2, k}^y-\bff_{i, j+1 / 2, k}^y\right]+ \\
        \frac{1}{\Delta z_k}
        \left[\bff_{i, j, k-1 / 2}^z-\bff_{i, j, k+1 / 2}^z\right] +
        \bs_{i, j, k}.
    \end{aligned}
\end{equation}

We use the HLLC approximate Riemann solver to compute the fluxes of the primitive variables across the cell faces.
The right and left states for the Riemann problem follow from a fifth-order accurate WENO reconstruction, which is robust to grid-scale phase and state discontinuities~\citep{bryngelson2019quantitative}.

\begin{figure}[th]
	\centering
	\includegraphics[scale=1]{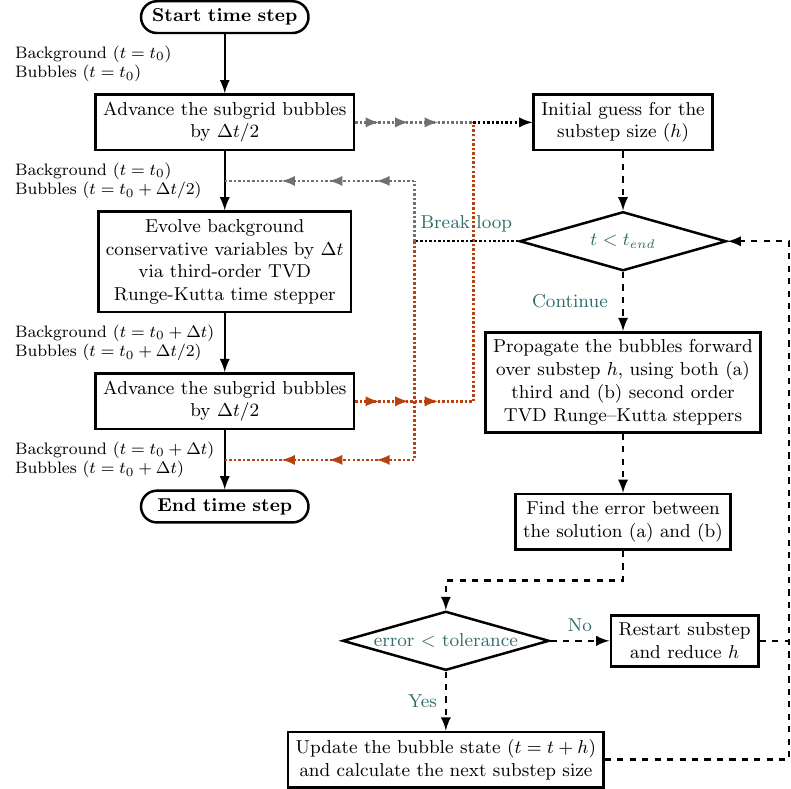}
	\caption{General schematic of the Strang splitting algorithm for time integration.}
	\label{fig:strangSplitting}
\end{figure}

To advance the conservative variables in time, we adopt a third-order total variation diminishing (TVD) Runge--Kutta time-stepping scheme~\cite{gottlieb1998}.
This method strikes an effective balance between numerical accuracy and stability, particularly in the presence of sharp gradients.
Unlike the relatively smooth evolution of the grid-resolved background flow, subgrid-scale bubbles often undergo transient and nonlinear dynamics, particularly during events such as bubble collapse and rebound.
These phenomena involve rapid variations in pressure and volume, necessitating significantly smaller time steps to maintain numerical stability and accurately capture the physics.
However, applying such fine time resolution uniformly across the bubbles and background flow would drastically increase computational cost, making the simulations infeasible.
To address this challenge, we employ a Strang splitting algorithm \cite{strang1968}, which decouples the time step of the background flow from that of the bubble dynamics.
\Cref{fig:strangSplitting} schematically illustrates the implementation of the Strang splitting method within our framework.
This approach enables us to treat the rapid, localized dynamics of subgrid bubbles independent of the slower evolution of the background flow.
In practice, the bubble equations are solved using smaller time steps nested within each coarser background time step.
To ensure accuracy and stability, the solver only advances each substep if the relative errors in the bubble radius $R$ and interface velocity $\dot{R}$ calculated using third- and second-order TVD Runge--Kutta steppers remain below a tolerance of $10^{-4}$. 
The algorithm alternates between advancing the states of the subgrid bubbles and the conservative variables of the flow field, ensuring consistent coupling between them.
By manipulating the timescale requirements of the two systems, this strategy enables the efficient simulation of complex multiscale phenomena without incurring prohibitive computational costs.

\section{Hardware acceleration strategy} \label{sec:accel_str}

The multiscale algorithm is hardware-accelerated using graphics processing units (GPUs).
The grid-resolved background flow follows the acceleration strategy developed by \citet{radhakrishnan2024}, which is implemented in MFC.
Building upon this framework, we extended the GPU acceleration to our previously described phase-averaged subgrid bubble models.
In the volume-averaged (Euler--Lagrange) formulation, the acceleration targets the evolution of discrete bubbles and the two-way coupling routines within the EL framework.
The ensemble-averaged (Euler--Euler) formulation focuses on accelerating the transport of bubble number density, void fraction, and averaged bubble dynamics.
This integrated approach enables full exploitation of GPU-based parallelism across both the micro- and macro-scale components of the simulation.

Once the initial conditions are established on the CPU, the state variables are transferred to the GPU, which subsequently performs the majority of the computational workload.
OpenACC directives are used to offload all computationally intensive tasks, ensuring that parallel regions and loop structures are automatically mapped to the GPU architecture.
After identifying independent loops and defining their levels of parallelization, the OpenACC runtime automatically selects optimal kernel configurations, maximizing performance and parallel efficiency.
This automated tuning process ensures that the algorithm is well-optimized for the target GPU hardware. A key advantage of this directive-based offloading approach is maintaining a unified codebase for both CPU and GPU execution.
Rather than maintaining separate implementations for different hardware platforms, the OpenACC directives allow the compiler to generate architecture-specific code at compile time via simple compiler flags.
By adhering to standard OpenACC syntax, the implementation remains portable across multiple computing environments, as these directives are supported by compilers such as NVHPC, GNU, and Cray (CCE).
This flexibility ensures compatibility with both NVIDIA and AMD GPUs.
In this study, GPU results were obtained using the NVHPC~24.1 SDK, and CPU-based runs used GNU~11.4.
No meaningful performance differences were observed for different compiler versions.
MFC also uses metaprogramming techniques enabled by the Fypp preprocessor to enhance GPU kernel performance~\cite{radhakrishnan2024}.
User-defined inputs are treated as compile-time constants, allowing the compiler to allocate fixed-size thread-local arrays and optimize memory access through register utilization.
This approach also removes conditional branching and redundant kernel duplication across spatial dimensions.
Additional optimizations are employed, which are described in full in \citet{radhakrishnan2024}.
These optimizations contribute to MFC's high computational performance and serve as the foundation for the acceleration strategy adopted in this work.

An example of the OpenACC kernel used to incorporate the influence of Lagrangian bubble volume within the Eulerian framework is shown in \cref{lst:OpenACC_dir_el}.
The \texttt{parallel loop} construct is augmented with a \texttt{gang vector} clause, which enables the compiler to automatically determine the optimal number of gangs and vectors, maximizing resource usage~\cite{chandrasekaran2017openacc}.
The outer loop iterates over all bubbles, and each bubble is subsequently processed through three nested loops spanning the spatial coordinates.
These loops are collapsed into a single memory-coalesced loop via the \texttt{loop collapse(3)} clause, improving loop-level parallelization across the three dimensions.
In cases where bubbles are located in proximity, multiple threads may attempt to update the same computational cell concurrently, leading to potential race conditions and data corruption.
To address this, OpenACC provides the \texttt{atomic update} clause, which enforces thread-safe access to memory.
When multiple threads update the same cell, their operations are serialized to ensure that all contributions are correctly accumulated without data loss or overwriting.
In our implementation, the \texttt{atomic update} clause is applied selectively to ensure accurate updates to the Eulerian grid induced by the bubbles, while preserving parallel efficiency.

\begin{figure}[tb]
\centering
\begin{minipage}[t]{0.48\textwidth}
\begin{lstlisting}[caption={EL Model: OpenACC directives to smear the bubble volume.}, label={lst:OpenACC_dir_el}, language=Fortran, escapeinside={(*@}{@*)}]
(*@\textcolor{rust}{!\$acc parallel loop gang vector}@*)
(*@\textcolor{rust}{!\$acc default(present) private(..)}@*)
do q = 1, nBubs
  cell = f_find_cell(q)
  stdev = f_kernel_width(q)
 
  (*@\textcolor{rust}{!\$acc loop collapse(3)}@*)
  do j = 0,delt; do k = 0,delt; do l = 0,delt
  ! Coordinate directions of local smearing
    aux(1) = cell(1) + j - delt
    aux(2) = cell(2) + k - delt
    aux(3) = cell(3) + l - delt
    call s_check_outside(aux)
    if (.not. outside) then
      call s_gaussian(stdev, fun)
    end if

    (*@\textcolor{rust}{!\$acc atomic update}@*)
    updVar%sf(aux) += fun*bubVolume
  end do; end do; end do
end do
(*@\textcolor{rust}{!\$acc end parallel loop}@*)
\end{lstlisting}
\end{minipage}
\hfill
\begin{minipage}[t]{0.48\textwidth}
\begin{lstlisting}[caption={EE Model: OpenACC directives to evolve the bubble state.}, label={lst:OpenACC_dir_ee}, language=Fortran, escapeinside={(*@}{@*)}]
(*@\textcolor{rust}{!\$acc parallel loop collapse(3)}@*)
(*@\textcolor{rust}{!\$acc gang vector default(present)}@*)
(*@\textcolor{rust}{!\$acc private(..)}@*)
do j = 0,Nx; do k = 0,Ny; do l = 0,Nz
! Loop over grid

  ! Bubble number density
  nbub = f_obtain_nbub()
   
  (*@\textcolor{rust}{!\$acc loop seq}@*)
  do q = 1,nBins
     bub = f_bubState(q)
     if (adap_dt) then !time advance
       ! Strang Split
       call s_advance(bub, R_loc, Rdot_loc)
       ! Update variables
       q(r(q))%sf(j,k,l)= nbub*R_loc
       q(v(q))%sf(j,k,l)= nbub*Rdot_loc
     end if
  end do
end do; end do; end do
(*@\textcolor{rust}{!\$acc end parallel loop}@*)
\end{lstlisting}
\end{minipage}
\end{figure}

Similarly, \cref{lst:OpenACC_dir_ee} presents the OpenACC kernel used to advance the time evolution of the averaged bubble state in the EE model.
As before, the \texttt{gang vector} clause allows the compiler to automatically select the optimal number of gangs and vectors to maximize utilization of the available GPU resources.
Three nested spatial loops are merged using the \texttt{loop collapse(3)} clause. In contrast, an additional \texttt{loop seq} clause is employed to iterate over the discrete bins of the log-normal probability distribution, which represents the initial bubble size distribution.
Within this loop, the Strang Splitting subroutine is invoked to apply the algorithm illustrated in \cref{fig:strangSplitting}.
Finally, the scalar fields in the Eulerian framework where the subgrid bubbles reside are updated accordingly.
In this step, the use of an \texttt{atomic update} clause is unnecessary, as each GPU thread operates on a unique entry (\texttt{j}, \texttt{k}, \texttt{l}, \texttt{q}) defined by its $x$-, $y$-, and $z$-coordinate indices and bin index, respectively.

\section{Results and discussion} \label{sec:results}

\subsection{Validation}

The accuracy of our volume-averaged multiscale solver is rigorously validated through two distinct test cases, each designed to assess its ability to accurately predict the dynamic behavior of oscillating and spherically collapsing bubbles.

\begin{figure}[th]
	\centering
    \includegraphics[scale=1]{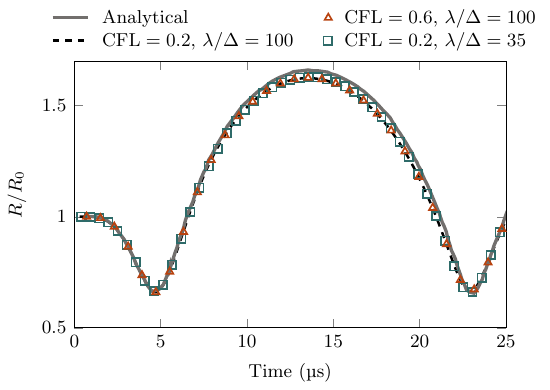}
	\caption{Evolution of an isolated bubble in response to a single cycle of a sinusoidal pressure wave using different CFL numbers and grid sizes.}
	\label{fig:el_oscillatingBub}
\end{figure}

In the first validation scenario, we consider an isolated gas bubble in water.
Its initial radius is \SI{50}{\micro\meter}, and is positioned at a fixed location ($0, 0, 0$) at the center of the computational domain.
This bubble is exposed to a planar sinusoidal acoustic wave with an amplitude of \SI{0.2}{\mega\pascal} and a frequency of \SI{150}{\kilo\hertz}.
To prevent acoustic reflections through the boundaries, we use the non-reflective boundary conditions described by \citet{thompson1987BC}.
The temporal evolution of the bubble radius, shown in \cref{fig:el_oscillatingBub}, is compared against the analytical solution of the Keller--Miksis equation, as reported by \citet{maeda2018eulerian}.
We performed a series of simulations to quantify the solver's variability with respect to the Courant–Friedrichs–Lewy (CFL) number and background grid size. 
Because the high-order TVD Runge–Kutta scheme is explicit, the numerical stability criteria require a CFL number below unity.
MFC can automatically adjust the time step to ensure that the CFL number remains below a user-defined maximum throughout the domain.
In our simulations, we observe negligible variability as the CFL number is increased from 0.2 to 0.6.
We also examined the influence of grid resolution while holding the CFL number fixed at 0.2.
The results again show negligible sensitivity, with comparable accuracy when resolving the acoustic wavelength using 35 or 100 cells.
Overall, this parametric study highlights the temporal and spatial robustness of the solver.
The numerical predictions exhibit excellent agreement with the analytical solution, with a maximum root-mean-squared error (RMSE) of \SI{2.76}{\percent}.

\begin{figure}[t]
	\centering
	\includegraphics[scale=1]{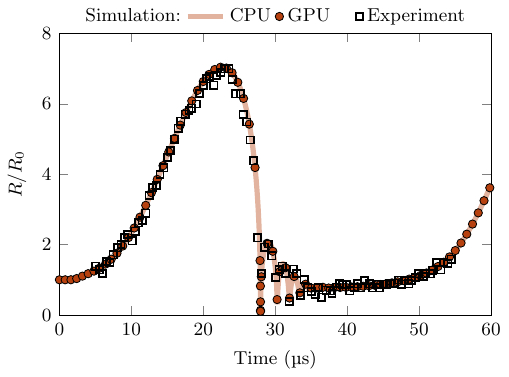}
	\caption{
        Radius evolution of the spherical collapse of an isolated bubble in response to a sinusoidal acoustic wave.
    }
	\label{fig:val_singleBubCollapse}
\end{figure}

In the second validation case, we replicate the experimental observations reported by \citet{ohl1999bubble}, who investigated the dynamics of a trapped bubble in a water--glycerine mixture undergoing spherical collapse.
The liquid host has a density of \SI{1000}{\kilogram\per\cubic\meter}, viscosity of \SI{0.006}{\newton\second\per\square\meter}, and the surface tension of the liquid-bubble interface is \SI{0.07}{\newton\per\meter}.
A single bubble with an initial radius of $R_0=\SI{8}{\micro\meter}$ is fixed in space and subjected to a sinusoidal acoustic wave with an amplitude of \SI{1.32}{\bar} and a frequency of \SI{21.4}{\kilo\hertz}.
The grid spacing is uniform across the domain with $\Delta = \SI{100}{\micro\meter}$, and the CFL number is kept at 0.2. 
For this particular simulation, we disabled the mass transfer model given in \cref{eqn:massVaporRate}. The reduced-order formulation developed by \citet{preston2007} assumes that the vapor pressure remains in equilibrium at the gas–liquid interface. Under strong bubble collapses, however, this assumption becomes less certain, as the rapid dynamics may prevent the interface from maintaining thermodynamic equilibrium.
Consequently, we chose not to include this model here to maintain consistency with the experimental observations.
The expansion part of the pressure wave makes the bubble grow, and we observe a maximum bubble radius of 7.04$R_0$, as depicted in \cref{fig:val_singleBubCollapse}.
Then, the compressive part of the wave provokes a sudden collapse of the bubble, followed by a rebound cycle.
The simulated bubble radius evolution during growth, collapse, and rebound cycles shows close agreement with the experimental measurements, yielding an RMSE of \SI{7.46}{\percent}.
Furthermore, the results obtained using GPU-based simulations are identical to those from CPU-based computations, confirming the correct implementation of GPU offloading and data-parallelism within our solver.

\begin{figure}[hb]
	\centering
	\includegraphics[scale=1]{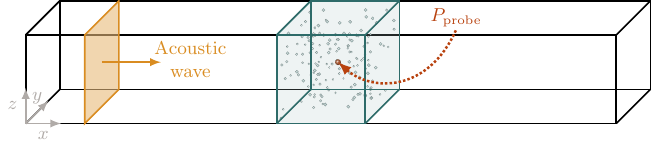}
	\caption{
        Schematic of the dilute bubble screen configuration (not to scale).
    }
	\label{fig:schematic_bubScreen}
\end{figure}

To evaluate the accuracy of the ensemble-averaged subgrid model, we simulate the interaction between a dilute bubble screen and a single sinusoidal planar acoustic wave.
Both the volume-averaged and ensemble-averaged subgrid models are employed to enable a systematic comparison.
Using our previously validated volume-averaged model as a reference, we test the performance of the ensemble-averaged formulation.
The computational domain is a square prism, shown in \cref{fig:schematic_bubScreen}, with spatial extents defined as $x \in [-20, 20]~\si{\milli\meter}$ and $y, z \in [-2.5, 2.5]~\si{\milli\meter}$.
The bubble distribution is cubic, centered at the origin $(0,0,0)$, extends \SI{5}{mm} along the $x$-direction, and spans the full extent of the domain in the $y$- and $z$-directions.
The screen consists of a cloud of bubbles with an initial void fraction of $\alpha_0 = 4 \times 10^{-5}$, ensuring the flow remains in the dilute regime, and all bubbles are initialized in their equilibrium states.

Two test cases are considered for the bubble size distribution: a monodisperse case and a polydisperse case.
In the monodisperse scenario, all bubbles have the same initial radius of \SI{10}{\micro\meter}.
In the polydisperse scenario, a more realistic distribution of bubble sizes is introduced by assigning radii according to a log-normal distribution, also centered at \SI{10}{\micro\meter}, with a log-normal shape parameter of $\sigma_p = 0.3$.
To excite the bubble screen, we impose a planar sinusoidal acoustic wave that is generated at $x = \SI{-7.5}{\milli\meter}$ and propagates in the positive $x$-direction toward the bubble cloud.
The wave has a frequency of \SI{300}{\kilo\hertz} and an amplitude of \SI{0.1}{\mega\pascal}.
The surrounding fluid is water, and the gas inside the bubbles is air, representing a typical two-phase system in acoustically active environments such as underwater acoustics or biomedical ultrasound.
The domain boundaries are configured with non-reflective boundary conditions to prevent acoustic reflections.

\begin{figure}[ht]
	\centering
	\includegraphics[scale=1]{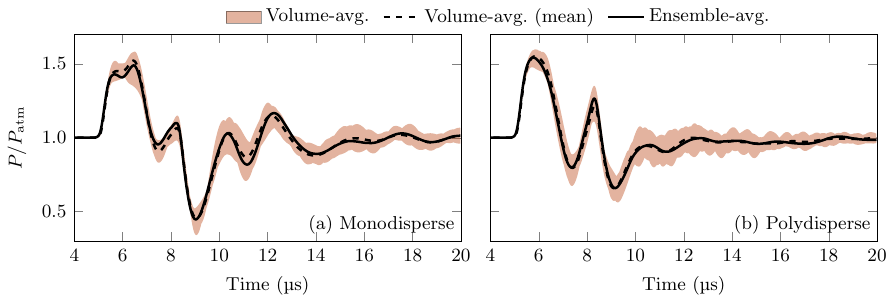}
	\caption{Pressure profiles at the origin of the bubble screen using the volume-averaged and ensemble-averaged subgrid models. (a) Monodisperse and (b) polydisperse bubble cloud.}
	\label{fig:val_bubScreen}
\end{figure}
A 3D Cartesian grid is used for spatial discretization.
The grid is uniformly spaced, with 400~cells in the $x$-direction and 50~cells in the $y$- and $z$-directions.
This configuration results in a uniform cell size of $\Delta_x = \Delta_y = \Delta_z = \SI{100}{\micro\meter}$.
The choice of grid resolution is guided by the need to resolve the acoustic wavelength with sufficient accuracy.
Specifically, the acoustic wave is resolved using 50 cells per wavelength, which is adequate to prevent numerical dissipation that could otherwise attenuate the wave during its propagation through the medium~\cite{gnanaskandan2019modeling}.
We maintain a CFL number of 0.2 throughout all simulations.

For simulations using the volume-averaged model, bubbles are randomly placed within the defined cubic bubble screen.
To account for statistical variations introduced by the random placement of bubbles and to extract a meaningful average behavior, we perform 40 independent simulations, which are sufficient for statistical convergence, as shown in a previous study~\citep{bryngelson2019quantitative}.
The mean inter-bubble distance for the given void fraction and bubble radius is approximately \SI{265}{\micro\meter}, which is more than twice the grid spacing.
This satisfies the requirement of \cref{eqn:model_inequality}.

In the ensemble-averaged modeling framework, the polydispersity of bubble sizes introduces additional complexity.
Instead of simulating the full spatial distribution of individual bubbles, the model resolves the average behavior across a spectrum of equilibrium bubble radii.
To capture this variability, the radius space is discretized into a finite number of bins, $N_{\mathrm{bin}}$, each representing a subset of the log-normal distribution.
\citet{bryngelson2019quantitative} conducted a parametric study to quantify the sensitivity of the model predictions for different $N_{\mathrm{bin}}$, showing that the error decreases as the number of bins increases.
For example, they report an \SI{8}{\percent} error for $\sigma_p = 0.3$ when using $N_{\mathrm{bin}} = 11$.
To reduce this error and enhance the model's predictive capability, we increase the resolution of the bubble size spectrum to $N_{\mathrm{bin}} = 21$ in our polydisperse simulations.
This choice provides a practical balance between accuracy and computational cost.

\Cref{fig:val_bubScreen} shows the pressure measurements at the center of the bubble cloud, $P_{\mathrm{probe}}(0,0,0)$, for both bubble distributions.
The shaded region represents the variability range observed across all Euler--Lagrange realizations.
In both monodisperse and polydisperse configurations, the pressure profiles computed using the ensemble-averaged model closely follow the mean behavior of the 40 volume-averaged simulations.
The RMSE error between the ensemble-averaged and volume-averaged models is \SI{2.10}{\percent} for the monodisperse case and \SI{1.53}{\percent} for the polydisperse case.
These results confirm the accuracy and validity of the ensemble-averaged subgrid model for dilute bubbly flows.

\subsection{Computational cost of the bubbles}

\begin{figure}[ht]
	\centering
	\includegraphics[scale=1]{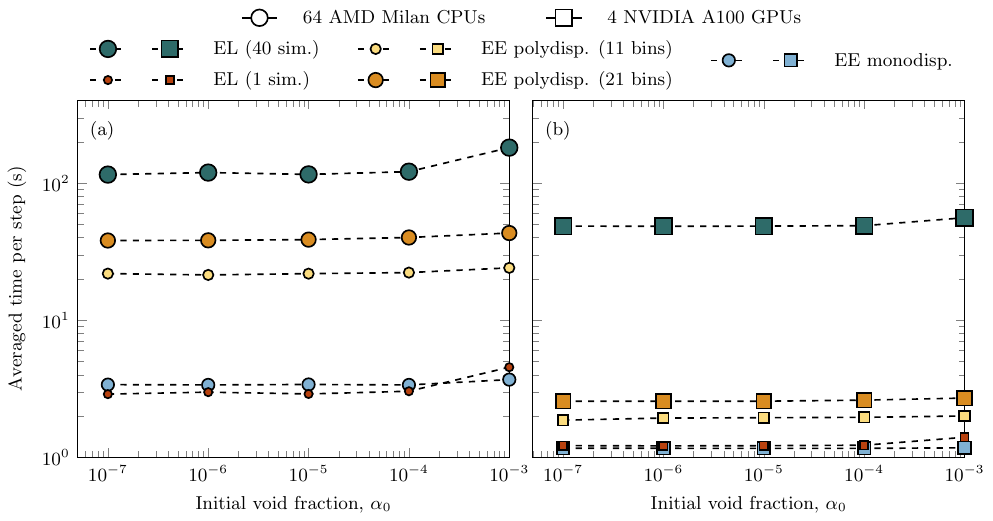}
	\caption{Computational cost of the volume-averaged (EL) and ensemble-averaged (EE) models on (a)~CPU cores and (b)~GPUs.}
	\label{fig:EL_bubbles_cost}
\end{figure}

In this section, we evaluate the computational expense associated with our phase-averaged bubble models and demonstrate the performance benefits of the hardware-accelerated solver utilizing GPUs.
We consider a representative bubble screen configuration with varying initial void fraction values, ranging from $\alpha_0 = 10^{-7}$ to $\alpha_0 = 10^{-3}$.
The computational domain is a cubic volume of water with dimensions $x, y, z \in [-7.5, 7.5]~\si{\milli\meter}$, within which a bubble cloud is initialized.
The bubble radii follow a log-normal distribution centered at $R_0 = \SI{10}{\micro\meter}$, with a shape parameter of $\sigma_p = 0$ for monodisperse and $\sigma_p = 0.3$ for polydisperse cases.
The bubbles are initialized out of equilibrium; a non-zero radial velocity $\dot{R_0} = \SI{0.015}{\meter\per\second}$ is prescribed at the gas--liquid interface, which in the polydisperse case varies linearly with bubble size.
This initialization promotes bubble oscillations in the absence of external excitation sources such as traveling pressure waves. The computational cost of the EL model primarily depends on the total number of oscillating discrete bubbles in the domain and is independent of the bubble size distribution.
Consequently, we restrict our EL tests to the monodisperse bubble cloud configuration.
In contrast, the cost of the ensemble-averaged model depends on the discretization of the bubble size distribution.
Specifically, the number of equations solved scales with the number of bins ($N_\mathrm{bin}$) used to represent the log-normal distribution of the bubble radii.
In our simulations, the EE monodisperse model corresponds to $N_\mathrm{bin}=1$, while the EE polydisperse cases employ $N_\mathrm{bin}=11$ and $N_\mathrm{bin}=21$.

The Eulerian background flow is discretized using a uniform grid of $\Delta_x = \Delta_y = \Delta_z = \SI{75}{\micro\meter}$.
To enable a fair comparison between the subgrid models, this background mesh is kept fixed across all simulations.
We use a constant time step of $\Delta t = \SI{0.03}{\micro\second}$, and each simulation runs for a total duration of \SI{1}{\micro\second} using 33 time steps.
After discarding the first three start-up time steps, we compute the averaged wall-clock time per step, as reported in \cref{fig:EL_bubbles_cost}. All simulations were executed on the NCSA~Delta supercomputer; detailed hardware specifications are available in \citet{a2023_delta}.
We use the 4-way NVIDIA A100 GPU compute nodes, which contain 4~NVIDIA A100 GPUs and 64~AMD Milan CPU cores per node.
Based on this, we use one node to run each of our simulations.
Thus, CPU-based simulations use 64~CPU cores and GPU simulations use 4~GPUs, each linked to one CPU core.
\Cref{fig:EL_bubbles_cost} summarizes the simulation outcomes.
The EE model demonstrates a clear advantage in computational efficiency, as it requires only one simulation to capture the mean dynamics of the dilute bubbly flow.
In contrast, the EL model requires multiple independent realizations (40 in this case, following \citet{bryngelson2019quantitative}) to achieve statistically converged results, due to the stochastic distribution of bubble locations, which significantly increases its overall cost.
The computational expense of the EE model scales with the number of bins.
The 21-bin configuration is the most expensive, followed by the 11-bin and monodisperse (1-bin) cases.
Notably, the performance of EE simulations remains largely unaffected by variations in void fraction across both CPU and GPU architectures.

\begin{table}[ht]
    \centering
    {\small
    \begin{tabular}{c| rr | rr}
         \multicolumn{1}{c}{} & \multicolumn{2}{c}{(a) 64 AMD Milan CPU cores} & \multicolumn{2}{c}{(b) 4 NVIDIA A100 GPUs}\\
        \toprule
        $\alpha_0$ & max(N.\ cells) & max(N.\ bubbles) & max(N.\ cells) & max(N.\ bubbles) \\
        \midrule
        $10^{-7}$ & 119458 &  6 & 1980050 & 81\\
        $10^{-6}$ & 119458 &  40 & 1980050 & 806\\
        $10^{-5}$ & 119458 &  255 & 1980050 & 8058\\
        $10^{-4}$ & 119458 &  2435 & 1980050 & 80573\\
        $10^{-3}$ & 119458 &  23891 & 1980050 & 805722 \\
        \bottomrule
    \end{tabular}
    }
    \caption{Euler--Lagrange simulation configurations, showing the maximum number of cells and discrete bubbles per processor on CPU and GPU architectures.}
    \label{tab:cost_cpu_gpu}
\end{table}

GPU acceleration substantially reduces computational time in all tested configurations.
We observe maximum speedups of 16x for the EE polydisperse case with 21 bins, 12x for 11 bins, 3.1x for the EE monodisperse case, and 3.2x for the EL model.
On CPUs, the EL model's cost increases significantly with void fraction, especially for $\alpha_0 = 1 \times 10^{-3}$, as a larger number of discrete bubbles intensifies the per-core workload.
Despite full parallelization, each CPU core handles bubble dynamics in a largely serialized fashion, in which its computational time scales directly with the number of bubbles assigned to it, which are shown in \cref{tab:cost_cpu_gpu}.
This leads to potential load imbalance if the bubble distribution is uneven.
Conversely, each NVIDIA~A100 GPU contains approximately 7K CUDA cores, enabling an additional layer of parallelization that accelerates bubble dynamics computations and mitigates imbalances.
The computational cost when performing a single simulation of the EL model is comparable to that of the EE monodisperse model. 

These computational cost assessments provide a comprehensive understanding of the expected performance of each subgrid model in both CPU and GPU architectures.
Beyond the performance metrics, the choice between the phase-averaged subgrid models depends largely on the underlying flow characteristics, the degree of spatial and temporal variability, and the level of statistical fidelity required for the analysis.
The EL model is particularly advantageous in configurations where resolving the spatial distribution and individual dynamics of discrete bubbles is required to represent or infer physical features.
Its Lagrangian framework naturally accommodates bubble--flow coupling, making it the most accurate choice when bubble motion or specific spatial distribution play a critical role.
Still, the requirement for multiple statistically independent realizations to obtain converged ensemble-averaged quantities significantly increases the computational expense of the EL approach, particularly for large void fractions or bubble counts.
In contrast, the EE model is more appropriate for statistically homogeneous bubbly flows where averaged properties are sufficient to describe the overall behavior.
Its formulation allows for a continuous representation of the dispersed phase through a set of discrete size bins, efficiently capturing both monodisperse and polydisperse distributions.
This discrete-bin treatment results in substantial reductions in computational cost while preserving the essential dynamics of the bubble population, including size evolution and interaction with the carrier fluid.
Another important consideration is memory usage, which becomes critical in large-scale simulations.
Among the models considered, the polydisperse EE approach is the most memory-intensive, particularly when a large number of bins is used to resolve the bubble size spectrum.
The total memory footprint scales with both the number of bins and the resolution of the Eulerian mesh, imposing practical constraints on simulations at high spatial resolutions or when a wide size distribution is represented.
Consequently, the balance between computational speed, memory requirements, and model fidelity must be carefully evaluated when selecting the appropriate subgrid model for a given problem.

The simulations discussed so far were carried out on a single compute node.
To assess their efficiency as the problem size changes, we proceed with a scaling study, described in the following section.

\subsection{Scaling study}

Here, we evaluate the scalability of the subgrid bubble models on both CPU- and GPU-based architectures through strong and weak scaling analyses.
The computational time is averaged over twenty time steps, which we find sufficient for the performance metrics to reach a steady state.
The physical configuration consists of a liquid domain filled with water serving as the base medium, within which a population of gas bubbles is dispersed at an initial void fraction of $\alpha_0 = 1 \times 10^{-3}$.
Consistent with our earlier study cases, the bubble sizes follow a log-normal distribution centered at \SI{10}{\micro\meter}, with $\sigma_p=0$ for monodisperse and $\sigma_p=0.3$ for polydisperse clouds. 
As in the previous study, the bubbles are initialized in a non-equilibrium state with an interface velocity of $\dot{R_0} = \SI{0.015}{\meter\per\second}$, which in the polydisperse case changes linearly with bubble size.
For the EE polydisperse scaling tests, we employ $N_\mathrm{bin}=3$, as increasing the number of bins significantly raises memory requirements and leads to out-of-memory issues during scaling studies.
For the EL tests, we focus on monodisperse bubble clouds, as in the previous section, with bubble positions randomly initialized to mimic realistic conditions.
The background mesh resolution is held constant across both weak and strong scaling studies, with $\Delta_x = \Delta_y = \Delta_z = \SI{80}{\micro\meter}$.

The scaling simulations were conducted on the NCSA Delta supercomputer~\cite{a2023_delta}.
CPU-based simulations were executed on Delta's CPU nodes, each equipped with dual AMD EPYC 7764 (Milan) processors, providing a total of 128 cores per node across 132~nodes.
GPU-based simulations were performed on Delta's 4-way NVIDIA A100 GPU compute nodes, comprising a total of 100~nodes.
Each GPU node includes four NVIDIA A100 GPUs and 64 AMD Milan CPU cores.
All Delta nodes are interconnected via the HPE/Cray Slingshot~11 high-performance interconnect, which supports data transfer rates of up to \SI{200}{\giga\bit\per\second} and enables high-bandwidth communication across the system.
In the GPU configuration, each GPU is directly coupled to a dedicated CPU core.

\subsubsection{Strong scaling}

\begin{figure}[!htb]
	\centering
    \includegraphics[scale=1]{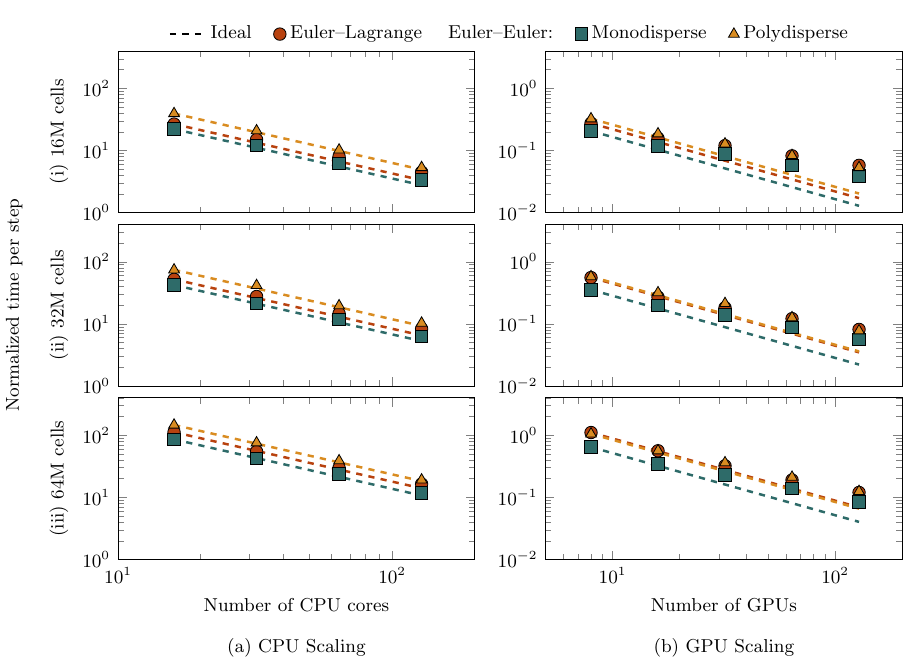}
	\caption{Strong scaling on (a) AMD CPUs and (b) NVIDIA A100 GPUs for different problem sizes: (i) 16M, (ii) 32M, (iii) 64M grid cells.}
	\label{fig:strong}
\end{figure}

In strong scaling, the total problem size is fixed, and performance is evaluated as the number of computational partitions increases.
Accordingly, the total number of Eulerian grid cells and, for the EL model, the total number of discrete bubbles, remains constant across simulations.
To comprehensively assess strong scaling behavior, three (\cref{fig:strong}~(i)--(iii)) increasingly large problem sizes were considered, consisting of 16 million, 32 million, and 64 million 3D Eulerian grid cells.
The corresponding EL simulations contained 2~million, 4~million, and 8~million discrete bubbles, respectively.

\Cref{fig:strong} presents the results of the strong scaling tests, and \cref{tab:efficiency_strong} reports the corresponding minimum parallel efficiencies.
For each problem size, the ideal scaling trend was defined using the average time per simulation step obtained at the smallest processor and GPU count, respectively.
Across both architectures, the computational cost trends are consistent with those observed in the previously presented subgrid cost assessment, where the EE polydisperse model exhibits the highest computational demand, followed by the single EL and EE monodisperse simulations.
On CPUs, the solver maintains high parallel efficiency as the problem size increases.
Even at the smallest problem size, efficiencies remain above \SI{74}{\percent} for the EL model and exceed \SI{82}{\percent} for both EE tests.
At 64M cells, the EE monodisperse and EE polydisperse cases reach efficiencies above \SI{90}{\percent}, indicating that the CPU implementation scales effectively and continues to benefit from additional cores.
For GPU-based tests, the solver closely follows the ideal scaling behavior up to intermediate configurations, after which the performance trend deviates as the number of GPUs increases, as shown in \cref{fig:strong}~(b).
Efficiency is lowest for the smallest problem and recovers as the problem size increases.
For the 64M problem size, efficiencies rise to \SI{58}{\percent} for EL and exceed \SI{47}{\percent} for the EE cases.
This upward trend suggests that the GPU solver achieves better scaling once each device has a sufficiently large computational workload.
The observed drop in efficiency for configurations with more than 16~GPUs is likely due to each GPU being underutilized, combined with the increasing cost of MPI communication.

\begin{table}[!t]
    \centering
    {\small
    \begin{tabular}{r| rrr | rrr}
         \multicolumn{1}{c}{} & \multicolumn{3}{c}{(a) AMD Milan CPU cores} & \multicolumn{3}{c}{(b) NVIDIA A100 GPUs}\\
        \toprule
         Grid Cells& 16M & 32M & 64M & 16M & 32M & 64M \\
        \midrule
        EL & \SI{74.21}{\percent} & \SI{78.64}{\percent} & \SI{85.31}{\percent} & \SI{29.69}{\percent} & \SI{42.70}{\percent} & \SI{57.68}{\percent}\\
        EE (monodisperse) & \SI{82.76}{\percent} & \SI{84.12}{\percent} & \SI{89.98}{\percent} & \SI{33.26}{\percent} & \SI{39.35}{\percent} & \SI{47.73}{\percent}\\
        EE (polydisperse) & \SI{92.43}{\percent} & \SI{89.19}{\percent} & \SI{94.02}{\percent} & \SI{38.13}{\percent} & \SI{46.52}{\percent} & \SI{52.77}{\percent}\\
        \bottomrule
    \end{tabular}
    }
    \caption{
        Minimum parallel efficiency of the strong scaling tests on CPU and GPU architectures.
        The grid cells are given in millions.
    }
    \label{tab:efficiency_strong}
\end{table}

These results demonstrate that the phase-averaged multiscale solver scales well on CPUs and efficiently utilizes the available computational resources.
They also suggest that GPUs are used most effectively when the hardware is fully occupied.

\subsubsection{Weak scaling}

\begin{figure}[ht]
	\centering
    \includegraphics[scale=1]{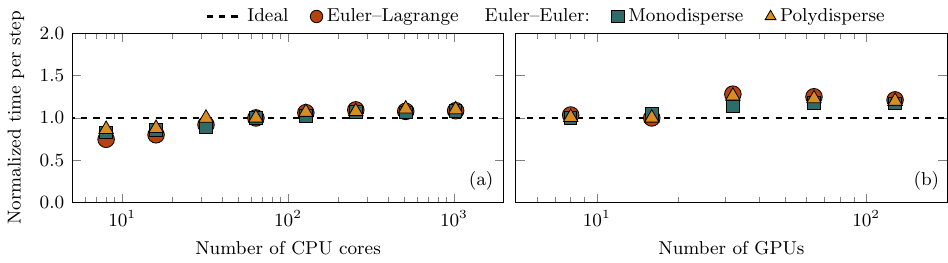}
	\caption{Weak scaling on (a) AMD CPUs and (b) NVIDIA A100 GPUs for Euler--Lagrange (EL) and Euler--Euler (EE) models as labeled.}
	\label{fig:weak}
\end{figure}

In weak scaling, the problem size assigned to each processor remains constant while the total number of compute devices increases.
Within the EE framework, the number of bins in the polydisperse tests is held constant, and the problem size is therefore only determined by the number of grid cells in the domain.
In contrast, for the EL model framework, maintaining a consistent problem size per processor requires fixing both the number of Eulerian grid cells, representing the continuous liquid phase, and the number of Lagrangian bubbles, representing the dispersed phase, assigned to each CPU or GPU.
To evaluate weak scaling behavior, we define a representative problem size in which each processor handles 100K 3D grid cells, and, in the EL model, an additional 20K discrete bubbles.

\Cref{fig:weak} presents the weak scaling performance of the solver.
The vertical axis indicates the average time per time step, normalized by the baseline cases using 64 CPU cores and 8 GPUs, respectively.
In an ideal scenario, a perfectly scalable solver would align with the dashed reference line.
The CPU-based simulations, depicted in \cref{fig:weak}~(a), exhibit excellent weak scaling, maintaining minimum parallel efficiencies of \SI{91.31}{\percent}, \SI{92.51}{\percent}, and \SI{90.29}{\percent} for the EL, EE monodisperse, and EE polydisperse models, respectively, for cases with 64 or more cores.
Each AMD socket comprises 64 cores organized into 4~NUMA (Non-Uniform Memory Access) domains.
Simulations executed with 8 or 16~cores are confined to a single NUMA domain, enabling low-latency access to local memory and improved performance.
When 32~cores are used, additional cross-NUMA communication introduces latency, and the 64~core case saturates the intra-socket communication bandwidth.
Beyond this point, performance stabilizes as all cores within the socket are fully utilized.
Conversely, the GPU-based simulations, shown in \cref{fig:weak}~(b), achieve minimum parallel efficiencies of \SI{78.05}{\percent}, \SI{84.81}{\percent}, and \SI{79.40}{\percent} for the EL, EE monodisperse, and EE polydisperse models, respectively.
Performance improves up to 16~GPUs, after which a gradual degradation is observed.
From 32 to 128~GPUs, the performance plateaus, likely due to inter-node communication reaching a saturation point.
The results confirm that the solver maintains strong parallel efficiency and scalability across both CPU and GPU architectures as computational resources increase, while keeping a constant problem size per compute device.

\section{Summary}\label{sec:summary}

We present a hardware-accelerated phase-averaged multiscale solver for acoustically driven bubbly flows.
The solver integrates both volume-averaged (EL) and ensemble-averaged (EE) models to capture the effects of the bubbles within a liquid host.
Its limitations are determined by the justified assumptions underlying each model.
In particular, we primarily assume bubble–bubble interaction occurs only through its effect on the liquid--bubble mixture, and the bubble radial oscillations obey the Keller--Miksis formulation.
The solver's physical accuracy was verified through three test cases.
The oscillation of a single bubble was compared with the analytical Keller--Miksis solution, achieving an RMSE of \SI{2.76}{\percent}.
We also considered the simulated spherical collapse of a gas bubble in a water-glycerine mixture which closely matched the experimental data from \citet{ohl1999bubble} with an RMSE of \SI{7.46}{\percent}, demonstrating the solver's ability to capture highly transient, nonlinear bubble dynamics.
Additionally, both CPU and GPU implementations produced identical results, confirming numerical consistency across architectures.
The EE model's ability to reproduce ensemble-scale pressure dynamics was verified by comparing its results with averaged data from 40~EL simulations.
For monodisperse and polydisperse bubble distributions, the RMSE between the EE and EL results was \SI{2.10}{\percent} and \SI{1.53}{\percent}, respectively, demonstrating that the EE formulation accurately represents the averaged behavior in statistically homogeneous bubbly systems.

The computational performance of both models was analyzed across varying void fractions.
The EL model's cost scales with the number of discrete bubbles, while the EE model's cost scales with the number of bubble size bins.
On CPUs, the EL model exhibited sharp cost increases at higher void fractions, resulting from serial bubble updates per core, which can lead to potential load imbalance.
In contrast, GPU-based EL simulations achieved up to 3.2 times speedups compared to 64-core AMD Milan CPU runs, due to the parallelism available via the NVIDIA A100 for bubble computations.
The EE model proved even more computationally efficient, as a single  simulation can represent ensemble dynamics.

On GPUs, the EE formulation achieved speedups of 3.1x (1~bin), 12x (11~bins), and 16x (21~bins) relative to 64-core CPU runs.
Although the 21-bin EE configuration was the most computationally demanding among the EE cases, it remained substantially cheaper than performing the 40 EL realizations required for ensemble convergence.
The computational cost of the EE model was largely insensitive to variations in void fraction, indicating robust scalability with respect to dispersed-phase volume fraction.
Memory usage, however, became a key consideration, particularly for the polydisperse EE model, where the cost scales with both the number of bins and Eulerian grid resolution.
For large-scale problems, the EE model's memory footprint can exceed that of the EL model despite shorter runtimes, emphasizing the need to balance accuracy, resolution, and hardware constraints when selecting between models.

Overall, GPU acceleration proved highly effective for both subgrid bubble models, reducing runtimes by more than an order of magnitude and enabling simulations that would otherwise be computationally prohibitive on CPU-only systems.
The demonstrated accuracy, scalability, and efficiency establish the solver as a robust tool for large-scale, high-fidelity simulations of multiphase systems relevant to biomedical ultrasound, underwater acoustics, and cavitating flows.

\section*{Acknowledgment}

We gratefully acknowledge the funding support from the U.~S. National Science Foundation (NSF) under the grants CBET 2301721 and CBET 2301709.
Additionally, this work used the Delta system at the National Center for Supercomputing Applications through allocation PHY230017 from the Advanced Cyberinfrastructure Coordination Ecosystem: Services \& Support (ACCESS) program, which is supported by National Science Foundation grants \#2138259, \#2138286, \#2138307, \#2137603, and \#2138296. 

\bibliographystyle{bibsty}
\bibliography{references.bib}

\end{document}